\documentclass[journal=jpcl,manuscript=letter]{achemso}

\usepackage[version=3]{mhchem} 
\usepackage[T1]{fontenc}       
\usepackage{color}

\author{Kuniyuki Miwa}
\affiliation[UCSD]{Department of Chemistry and Biochemistry, University of California San Diego, La Jolla, CA 92034, USA}
\author{Amin Morteza Najarian}
\affiliation[University of Alberta]{Department of Chemistry, University of Alberta, Edmonton, Canada}
\author{Richard L. McCreery}
\affiliation[University of Alberta]{Department of Chemistry, University of Alberta, Edmonton, Canada}
\author{Michael Galperin}
\email{migalperin@ucsd.edu}
\phone{+1-858-246-0511}
\affiliation[UCSD]{Department of Chemistry and Biochemistry, University of California San Diego, La Jolla, CA 92034, USA}

\title[Photocurrent in Junctions]
  {Hubbard NEGF Analysis of Photocurrent in Nitroazobenzene Molecular Junction}

\abbreviations{NAB, NEGF, DFT}

\begin{document}

\begin{tocentry}
\vspace*{0.2in}
\hspace*{0.7in}
\includegraphics{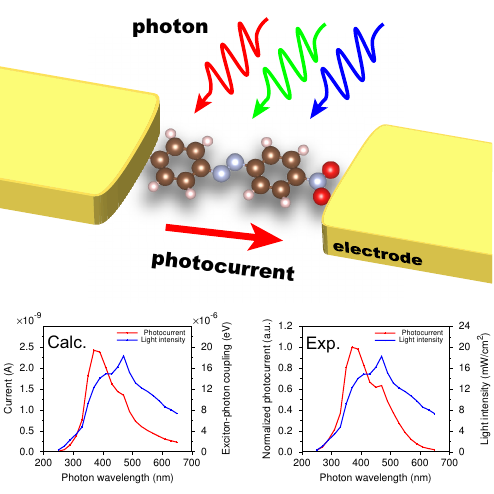}
\end{tocentry}

\begin{abstract}
  We present combined experimental and theoretical study of photo-induced current
  in molecular junctions consisting of monolayers of nitroazobenzene oligomers chemisorbed 
  on carbon surfaces and illuminated by UV-Vis light through a transparent electrode. 
  Experimentally observed dependence of the photocurrent on light frequency, temperature 
  and monolayer thickness is analyzed within first principles simulations employing 
  the Hubbard NEGF diagrammatic technique. We reproduce qualitatively correct behavior 
  and discuss mechanisms leading to characteristic behavior of dark and photo-induced currents 
  in response to changes in bias, frequency of radiation, temperature and thickness of molecular 
  layer.   
 \end{abstract}

The interaction of light with molecules is an important field of research due to its ability
to provide information on molecular structure and dynamics, and to serve as a control tool
for intra-molecular processes. 
Development of nano-fabrication and optical techniques at nanoscale led to tremendous
progress in ability to detect and manipulate molecules on surfaces and in junctions.
The main signal reported in the literature for devices consisting of
molecules attached to macroscopic leads for a long time was current-voltage
(conductance-voltage) characteristics \cite{AviramRatnerCPL74,NitzanRatnerScience03,LindsayRatnerAdvMat07,TaoNatNano06,GalperinRatnerNitzanTroisiScience08,MGRatnerNitzanJPCM07}.
Later, standard junction spectroscopies (such as resonant \cite{ZhitenevMengBaoPRL02,DekkerPRA77} and off-resonant \cite{Bayman1981,HoPRL00,ReedNL04,RuitenbeekNature02,Agrait2002}
inelastic electron tunneling spectroscopy)
were complemented by probing molecular conduction junctions by optical means \cite{ZhangNL07,HoPRB08,AllaraNL10,NatelsonNL08,CheshnovskySelzerNatNano08,NatelsonNatNano11,KimJCP11,SelzerChemSocRev11,ApkarianACSNano12,DongHouNature13}.
For single-molecule junctions the latter is possible only by local electromagnetic
field enhancement associated with plasmon excitations \cite{NatelsonNL07,SelzerNL11}.
Recent developments include observation and optical control of current \cite{SelzerJPCL13,ApkarianJCP13,NatelsonACSPhotonics15}
noise \cite{BerndtPRL10,BerndtPRL12}, and energy transfer~\cite{MiwaNature16,MiwaPRL17}.
Time-dependent and transient effects in junctions observed with optical means \cite{VanDuyneJACS13,ApkarianNL13,ApkarianNatPhoton14,SelzerJACS14},
tip-enhanced Raman spectroscopy \cite{VanDuyneACSNano13,VanDuyneJPCL14,VanDuyneJPCL14_2,VanDuyneNL15},
pump-probe spectroscopy in nanojunctions \cite{LothAPL13,ApkarianPotmaAPL15},
and reporting quantum interference effects \cite{VanDuyneNL12,ApkarianACSNano14}
are also among recent developments.
Optical spectroscopy yields a way to estimate heating in current carrying junction
\cite{CheshnovskySelzerNatNano08,NatelsonNatNano11,NatelsonNL14,VanDuyneJPCC15}
Recently, multidimensional spectroscopy measurements in the presence of current (although
not yet in junctions) were reported in the literature.\cite{CundiffOptExpres13,MarcusNatComm14,BittnerSciRep16}.
Optical effects have been also reported in large-area molecular junctions, including internal photoemission~\cite{RLM1},
optical modulation of conductance~\cite{RLM2}, light emission~\cite{RLM3,RLM4,RLM5,RLM6}, 
and photocurrents induced by light absorption~\cite{RLM7,RLM8}. 
Experimental capabilities to study radiation field interaction with molecular conducting
junctions gave rise to new branch of nanoscale research - {\em molecular optoelectronics} \cite{MGANPCCP12,MGChemSocRev17}

Theoretically, challenges in describing optical response in molecular junctions include 
necessity to account for open character of the system which requires simultaneous treatment of
optical transitions in the molecule and electron transfer between molecule and contacts.
Single particle language utilized in majority of ab initio studies in molecular electronics 
usually in the framework of the nonequilibrium Green functions (NEGF) 
makes it inconvenient to account for differences between transport and optical gaps in 
junctions. A possible alternative is utilization of molecular many-body states as a basis of 
consideration. In junctions, such consideration requires utilization of one of many-body flavors
of  the NEGF. In particular, recently introduced by us diagrammatic technique for the
Hubbard NEGF~\cite{ChenOchoaMGJCP17} yields a stable and accurate 
many-body method~\cite{MiwaChenMGSciRep17} 
conveniently suited for description of optoelectronic devices~\cite{MGChemSocRev17}.

Here, we apply the methodology to perform first principles simulation of optical response 
of recently reported large area  nitroazobenzene molecular junctions. 
Illumination by UV-Vis light induces major changes in current-voltage response, with orders of magnitude changes 
in conductance and distinct bias dependence compared to the dark behavior~\cite{RLM7}.
Below after introducing theoretical model
and calculational procedure, we present results of first principles simulations and compare them
with corresponding experimental data. We then discuss possible mechanisms for
changes in dark and photo-induced currents in response to bias, temperature, light frequency, 
and thickness of the molecular layer.

\begin{figure}[htbp]
\includegraphics[width=0.8\linewidth]{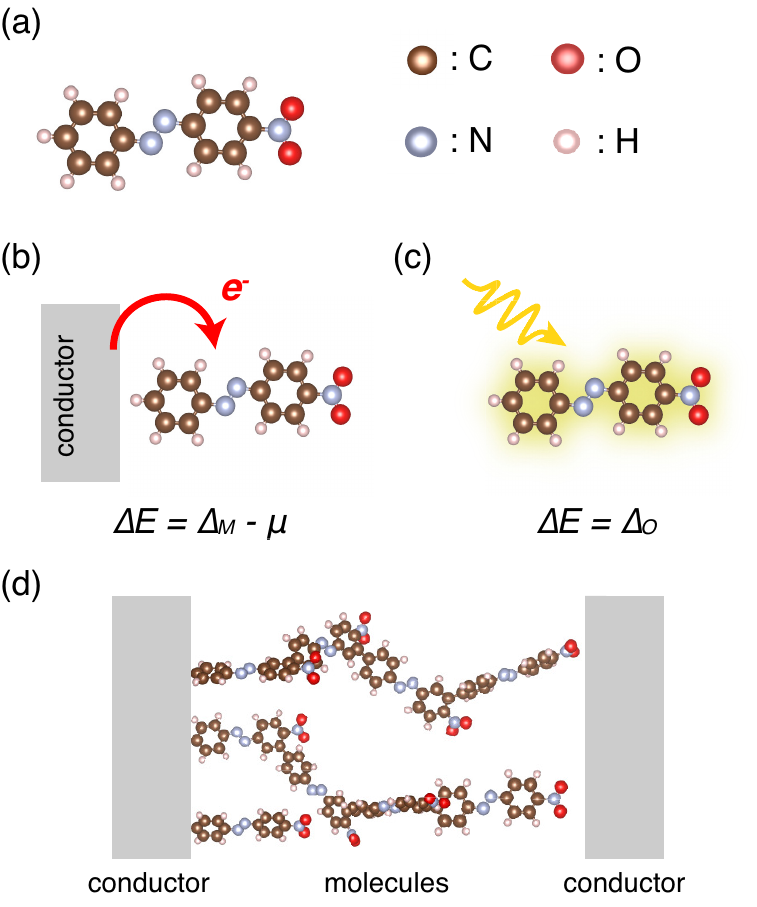}
\caption{\label{fig1}
Nitroazobenzene molecular junction. Shown are (a) molecule and 
sketches of (b) electron transfer process between molecule and contact
and (c) optical excitation by external radiation field. Panel (d) sketches molecular chains in the junction.
}
 \end{figure}

We consider a nitroazobenzene molecule (see Fig.~\ref{fig1}a), $M$, 
bridging two metal electrodes, $L$ and $R$, and subjected
to an external laser radiation, $rad$. The Hamiltonian of the junction is
\begin{equation}
\hat H = \hat H_M+ \sum_{B=L,R,rad}\bigg(\hat H_B+\hat V_B\bigg)
\end{equation}
Here $H_M$ and $H_B$ are respectively Hamiltonians of molecule and baths,
and $V_B$ introduces coupling between them.
We represent molecular Hamiltonian $\hat H_M$ as a tight-binding chain of $N$ molecular units (see Fig.~\ref{fig1}d)
\begin{equation}
\label{HM}
 \hat H_M=\sum_{n=1}^N\hat H_M^{(n)} + \sum_{n=1}^{N-1}\bigg(\hat V_M^{(n,n+1)} + H.c.\bigg)
\end{equation} 
For simplicity, we consider all the units to be identical.
We represent molecular unit Hamiltonian $\hat H_M^{(n)}$ in terms of many-body states 
$\lvert S_n\rangle$ of the unit. In particular, we consider ground, $\lvert N_g\rangle$,
and excited, $\lvert N_x\rangle$, states of
neutral molecular unit as well as ground states of anion $\lvert A_g\rangle$ and cation $\lvert C_g\rangle$. 
First-principle calculations (see Supporting Information for details) yield energies of 
the states $E_{S_n}$, and explicit form of molecular Hamiltonian is
\begin{equation}
\label{HMn}
\hat H_M^{(n)}=\sum_{S_n\in M_n} E_{S_n} \hat X_{S_nS_n}
\end{equation}
where $M_n$ is molecule $n$ in the chain and $\hat X_{S_nS_n}\equiv\lvert S_n\rangle\langle S_n\rvert$ 
is the Hubbard (or projection) operator.

Contacts are modeled as reservoirs of free electrons each at its own equilibrium
\begin{equation}
\hat H_K = \sum_{k\in K} \varepsilon_k \hat c_k^\dagger\hat c_k \qquad (K=L,R)
\end{equation}
and radiation field is described as continuum of modes with one mode corresponding
to laser frequency populated while all other modes empty
\begin{equation}
\hat H_{rad} = \sum_\alpha \omega_\alpha \hat a_\alpha^\dagger\hat a_\alpha
\end{equation} 
Here $\hat c_k^\dagger$ ($\hat c_k$) and $\hat a_\alpha^\dagger$ ($\hat a_\alpha$)
creates (annihilates) electron is state $k$ of the contacts and mode $\alpha$ of the field,
respectively. 

Within each molecular unit we consider four electron transitions (see Fig.~\ref{fig1}b)
$ET=A_g\to N_g$, $A_g\to N_x$, $N_g\to C_g$, and $N_x\to C_g$
($\Delta_{ET}=E_{A_g}-E_{N_g}$, $E_{A_g}-E_{N_x}$, $E_{N_g}-E_{C_g}$, 
$E_{N_x}-E_{C_g}$) and one optical transition (see Fig.~\ref{fig1}c) $OT=N_x\to N_g$
($\Delta_{OT}=E_{N_x}-E_{N_g}$). 
So that, electron transfer between the units is 
\begin{equation}
 \label{VMn}
 \hat V_M^{(n,n+1)} =  \sum_{{ET}_n\in M_n}\sum_{{ET}_{n+1}\in M_{n+1}}
 \bigg( t_{{ET}_n,{ET}_{n+1}}\hat X_{{ET}_n}^\dagger\hat X_{{ET}_{n+1}}
 + H.c.\bigg)
\end{equation}
where ${ET}_n$ are electron transfer transitions in $M_n$.
First molecule of the chain, $n=1$, is coupled to contact $L$, while last, $n=N$, couples the chain to contact $R$
\begin{align}
\hat V_K &= \sum_{\ell\in L} \sum_{ET_1\in M_1} 
\bigg( V_{\ell,ET_1}\,\hat c_\ell^\dagger\,\hat X_{ET_1} + H.c. \bigg)
\\ &+ \sum_{r\in R} \sum_{ET_N\in M_N} 
\bigg( V_{r,ET_N}\,\hat c_r^\dagger\,\hat X_{ET_N} + H.c. \bigg)
\\
\hat V_{rad} &= \sum_{n=1}^{N}\sum_{\alpha} \sum_{OT_n\in M_n}
\bigg(U_{\alpha,OT_n}\, \hat a_\alpha^\dagger\, \hat X_{OT_n} + H.c. \bigg)
\end{align}

Our central object of interest is current through the junction caused by either applied bias
($V_{sd}=\mu_L-\mu_R$), or laser field, or both - correspondingly, dark, optical, and total fluxes.
Current is introduced as rate of change of the population on the contacts, 
$I_K=d_t\sum_{k\in K} \langle \hat c_k^\dagger(t)\hat c_k(t)\rangle$, and at steady-state
considered here currents across $L$ and $R$ junction interfaces are equal (with opposite sign),
$I_L=-I_R$. Explicit current expression is given by the celebrated Meir-Wingreen 
formula~\cite{HaugJauho_2008}
\begin{equation}
 I_K = \frac{e}{\hbar} \mbox{Tr}\int\frac{dE}{2\pi} 
 \bigg(\sigma_K^{<}(E)\,\mathbf{G}^{>}(E) - \sigma_K^{>}(E)\,\mathbf{G}^{<}(E) \bigg)
\end{equation}
where trace is over electronic transitions $ET$ and
$\sigma_K^{<\, (>)}(E)$ and $\mathbf{G}^{<\, (>)}(E)$ are the Fourier transforms of the lesser (greater)
projections of electronic self-energy due to coupling to contact $K$ and Hubbard Green's
function, respectively. On the Keldysh contour the correlation functions are defined as
\begin{align}
 [\sigma_K(\tau_1,\tau_2)]_{{ET}_i,{ET}_j} &=
 \sum_{k\in K} V_{{ET}_i,k}\, g_k(\tau_1,\tau_2)\, V_{k,{ET}_j}
 \\
 G_{{ET}_i,{ET}_j}(\tau_1,\tau_2) &=
 -i\langle T_c\, \hat X_{{ET}_i}(\tau_1)\,\hat X_{{ET}_j}(\tau_2) \rangle
\end{align}
Here $\tau_{1,2}$ are the contour variables, $T_c$ is the contour ordering operator,
and $g_k(\tau_1,\tau_2)\equiv -i\langle T_c\,\hat c_k(\tau_1)\,\hat c_k^\dagger(\tau_2)\rangle$
is the Green's function for free electron in state $k$.

While explicit expressions for projections of the self-energy are known, 
Green's function has to be evaluated by solving
a modified Dyson equation on the Keldysh contour.
Because the Hubbard GF both depends and defines its self-energies, 
one has to solve the Dyson equation self-consistently until convergence
(see Supporting Information and Refs.~\cite{ChenOchoaMGJCP17,MiwaChenMGSciRep17} 
for details).

\begin{figure}[htbp]
\includegraphics[width=\linewidth]{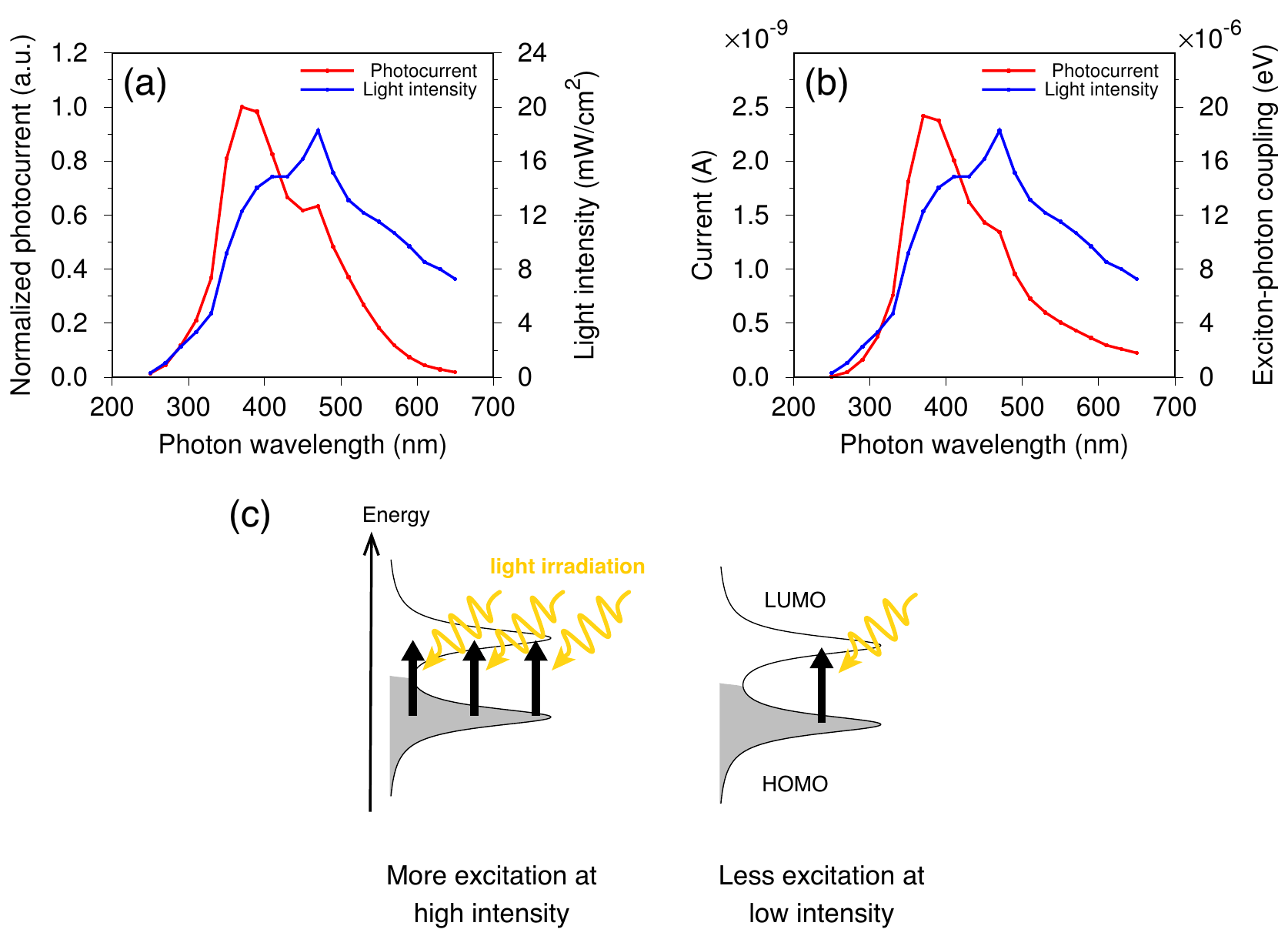}
\caption{\label{fig2}
Photocurrent (red line) and light intensity (blue line) as function of radiation field
wavelength. Shown are (a) experimental data and (b) theoretical photocurrent resulting from first-principles
simulations within the Hubbard NEGF. Panel (c) shows sketch of the mechanism.
See text for parameters.
}
 \end{figure}

Unless stated otherwise, simulations are performed at temperature
$T=300$~K. {\em Ab initio} (TD)DFT calculations of an isolated molecular unit yield electronic transitions
$\Delta_{N_gC_g}=-3.5$~eV, $\Delta_{N_xC_g}=-0.08$~eV,
$\Delta_{A_gN_g}=-0.11$~eV, $\Delta_{A_gN_x}=3.31$~eV, 
while optical gap is $\Delta_{OT}=3.41$~eV. 
Strength of molecule-contacts coupling is characterized by escape rate matrix
\begin{equation}
 \Gamma^K_{{ET}_i,{ET}_j}(E)\equiv 2\pi\sum_{k\in K}
 V_{{ET}_i,k}V_{k,{ET}_j}\delta(E-\varepsilon_k)
\end{equation}
which within assumed here wide band approximation does not depend on energy.
Simulations are performed for $\Gamma^L_{{ET}_i,{ET}_j}=0.9$~eV
and $\Gamma^R_{{ET}_i,{ET}_j}=0.3$~eV
for ${ET}_{i,j}\in\{A_g\to N_x,N_g\to C_g\}$
and $\Gamma^L_{{ET}_i,{ET}_j}=0.6$~eV
and $\Gamma^R_{{ET}_i,{ET}_j}=0.6$~eV
for ${ET}_{i,j}\in\{A_g\to N_g,N_x\to C_g\}$,
which are taken to reproduce experimental data
(see also Supporting Information).
Radiation field is assumed to be coupled to individual molecular units only.
Strength of the coupling to radiation field is characterized by dissipation rate
\begin{equation}
 \gamma_{OT_n}(\omega)\equiv 2\pi\sum_\alpha 
 U_{OT_n,\alpha} U_{\alpha,OT_n}\,\delta(\omega-\omega_\alpha)
\end{equation}
Its frequency dependence is taken from experimental data (see Fig.~\ref{fig2}a)
with value at molecular resonance, $\omega=\Delta_{OT_n}$, chosen $1.4\times 10^{-4}$~eV.
Fermi energy is taken as an origin, $E_F=0$, and bias is applied symmetrically:
$\mu_L=E_F+|e|V_{sd}/2$ and $\mu_R=E_F-|e|V_{sd}/2$.
Simulations were performed on energy grid spanning range from
$-16.384$ to $16.384$~eV with step $2$~meV.
Convergence was assumed to be reached when populations of the many-body states
at subsequent steps of the self-consistent procedure differ by less than $5\times 10^{-4}$.

We first consider photo-induced flux in the absence of bias. 
Here, we employ single molecular unit model, $N=1$, in the analysis. Optical excitation promotes
electronic population from ground to excited state, then electron either relaxes back or
escapes into contacts. Asymmetric coupling of excited state leads to appearance of
directed flux even in absence of bias. Figure~\ref{fig2} compares
experimental data (panel a) with  first principles calculation (panel b). 
Experimental data for light intensity (blue line) 
was used as an input in the calculations. As expected, photocurrent shows maximum
at frequency corresponding to optical transition $\Delta_{OT}$ (panel c). 
Such light-induced transport was discussed in theoretical literature within simple 
model considerations~\cite{GalperinNitzanPRL05}.
Shoulder observed at longer wavelengths is associated with peak in light intensity
at this frequency. One sees qualitative correspondence between experimental data
and theoretical simulations.

\begin{figure}[htbp]
\includegraphics[width=\linewidth]{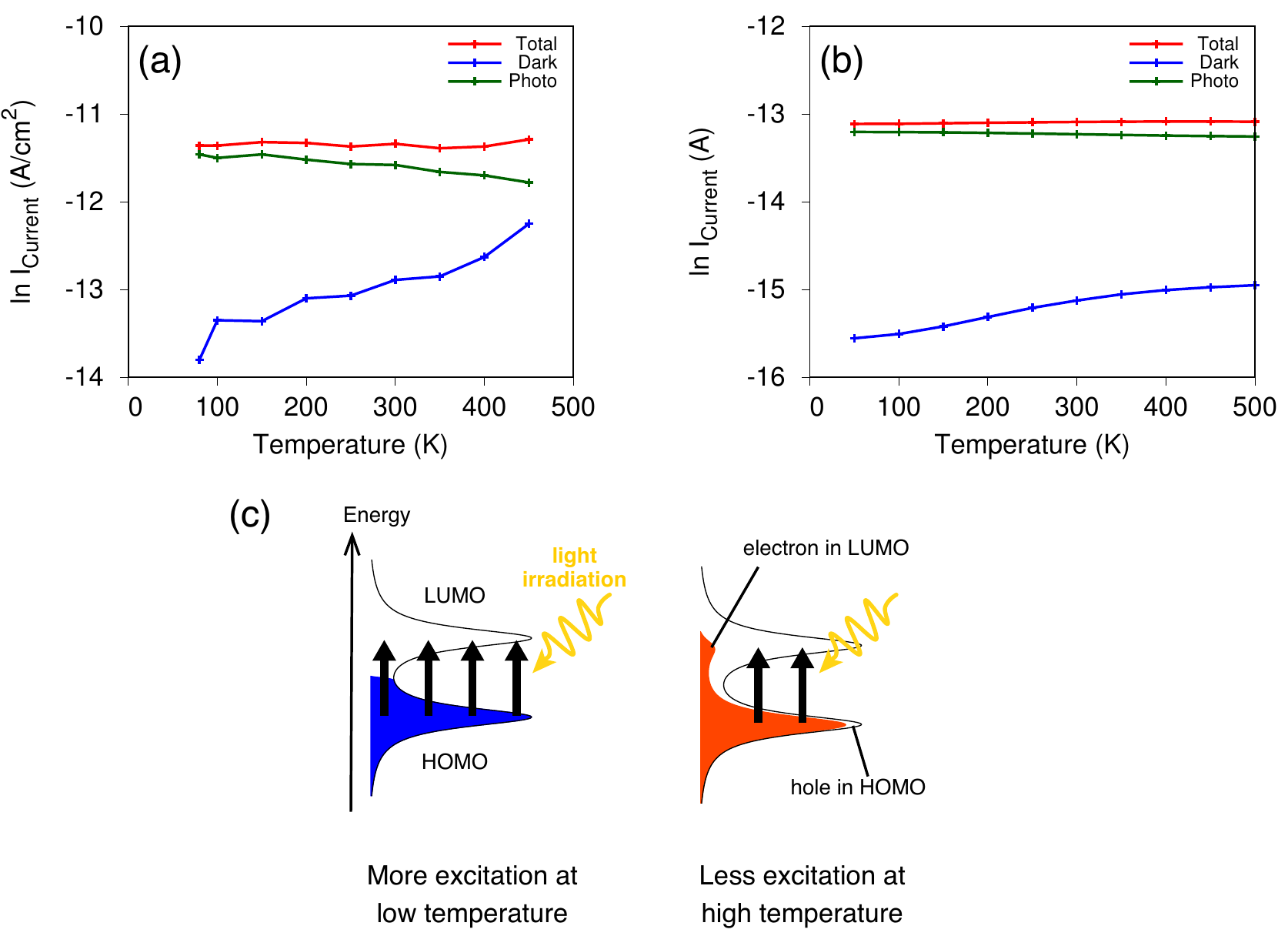}
\caption{\label{fig3}
Dark (blue line), photo-induced (green line), and total (red line) currents vs.
temperature of the junction.
Shown are (a) experimental data and (b) results of first-principles
simulations within the Hubbard NEGF. 
Panel (c) shows sketch of the mechanism.
See text for parameters.
}
 \end{figure}

We now turn to consideration of temperature dependence of current.
Also here single molecular unit model, $N=1$, is enough to describe the observed physics
Here junction is subjected to bias of $V_{sd}=0.02$~eV and 
measurements are performed in the absence of external radiation 
(dark current) and under radiation of $\lambda=380$~nm.
With temperature increase the experimental data (see Fig.~\ref{fig3}a) show an increase in dark current.
At the same time, photo-current (defined as difference between total and dark fluxes)
decreases. We note that in the experiment dark current
is caused by off-resonant tunneling, while radiation transfers electronic population from 
ground to excited state in near resonance conditions. Thus plausible mechanism can be 
suggested based on the smearing of Fermi distributions in the contacts with
temperature increase. In absence of external radiation extended tail of Fermi distribution
at higher temperatures yields more electronic population closer to molecular resonance,
which naturally leads to increase of electron flux. On the contrary, optical flux already being
at resonance mostly depends on available electronic population at ground state. 
Smearing of Fermi distribution slightly diminishes the latter thus resulting in decrease of 
photo-induced current (see Fig.~\ref{fig3}c). First principles simulations based on the proposed
mechanism (see Fig.~\ref{fig3}b) demonstrate qualitative correspondence with experimental data.
We note 
that possible additional factor (not included into the model) decreasing photocurrent with temperature 
is scattering within the monolayer~\cite{RLM7}.

\begin{figure}[htbp]
\includegraphics[width=\linewidth]{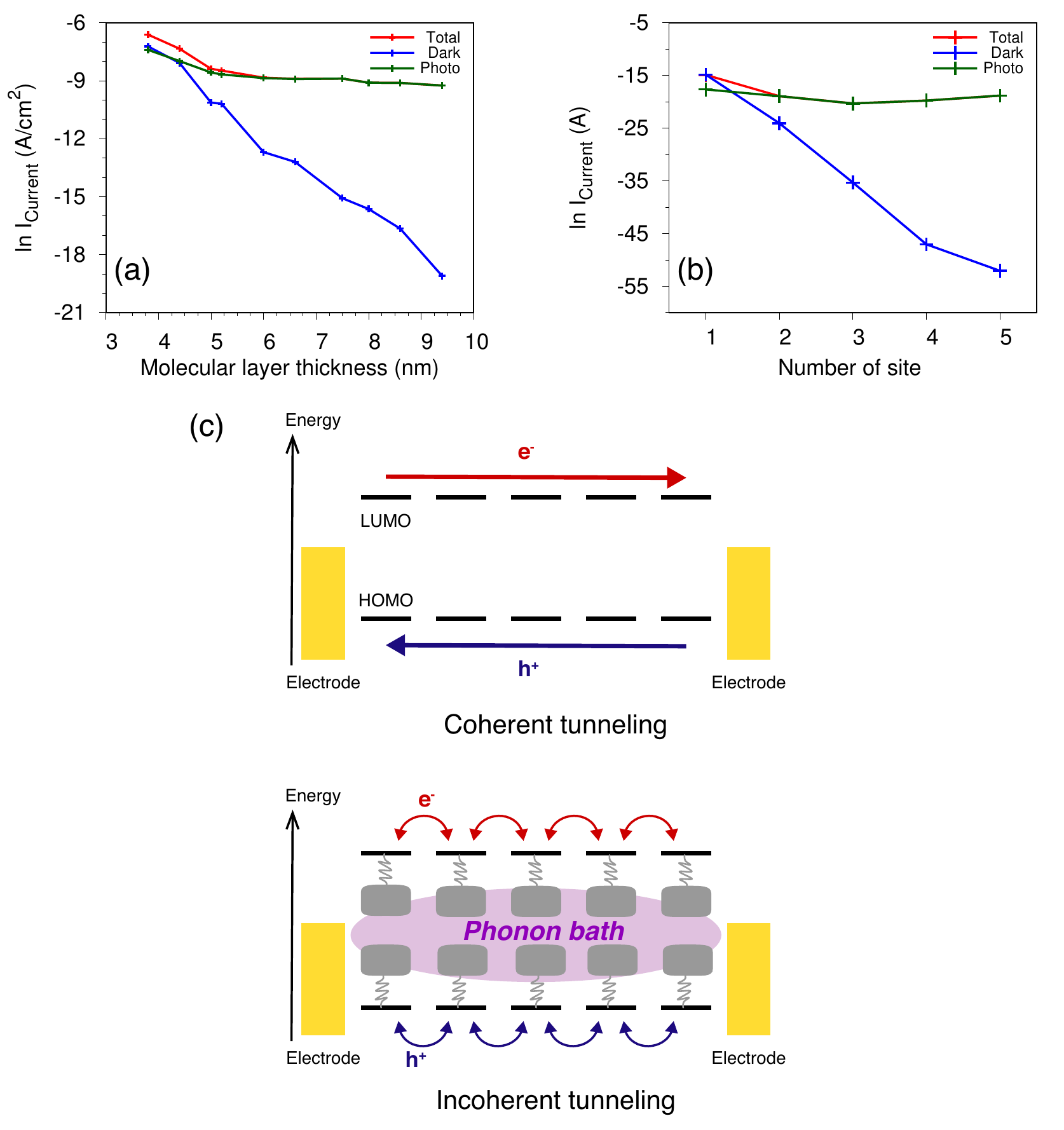}
\caption{\label{fig4}
Dark (blue line), photo-induced (green line), and total (red line) currents vs.
thickness of the molecular layer.
Shown are (a) experimental data and (b) results of first-principles
simulations within the Hubbard NEGF. 
Panel (c) shows sketch of the mechanism.
Simulations are performed under constant bias of $V_{sd}=0.02$~eV and external field illumination at wavelength $\lambda=380$~nm.
See text for other parameters.
}
 \end{figure}

Finally, we consider dependence of photocurrent on thickness of molecular layer.
The consideration requires performing calculations for molecular chains of different lengths.
In the analysis below we consider chains from $N=1$ to $N=5$ units.
While dark current demonstrates exponential dependence on the layer size
characteristic of tunneling, photocurrent practically does not depend on the thickness
(see Fig.~\ref{fig4}a).
Such insensitivity is expected for hopping transport, and interplay between the two
modes of behavior was discussed in the literature as result of competition
between tunneling and thermally activated electron transfer via molecular 
bridges~\cite{SegalNitzanJPCB00,ho_choi_electrical_2008,FrisbieJACS10}. We suggest that the same mechanism 
is behind observed length dependence also in transport with a difference that 
optical excitation in present case plays the role of thermal activation in previous study
(see Fig.~\ref{fig4}c).

Hopping character of electron transport is caused by local dephasing
(e.g., due to interaction with low frequency vibrations of nearby molecules in the layer),
which is usually modeled by introducing Buttiker probes.
Following Ref.~\cite{WhitePeskinMGPRB13} we represent Buttiker probes by attaching
set of oscillators to each many-body state of molecules
\begin{equation}
 \hat H_P = \sum_{n=1}^N \sum_{\beta_n}\sum_{S_n\in M_n}\bigg(
 \omega_{S_n} \hat b_{\beta_n}^\dagger\hat b_{\beta_n}
 + B_{\beta_n} \big(\hat b_{\beta_n}+\hat b_{\beta_n}^\dagger\big)\hat X_{S_nS_n}\bigg)
\end{equation}
and considering a limit of $\omega_{S_n}\to 0$ to restrict the effect of oscillators 
to pure dephasing. 

We treat both inter-molecule coupling and interactions with Buttiker probes
within second order of diagrammatic perturbation theory for 
the Hubbard NEGF~\cite{ChenOchoaMGJCP17}. This leads to appearance of two additional
self-energies in the self-consistent scheme (see Supporting Information for details).
The two interactions are characterized by inter-atomic electron tunneling parameters
$t_{\mathcal{M}_1\mathcal{M}_2}$ and dephasing rate
\begin{equation}
\gamma^{S_n}_P(\omega) = 2\pi \sum_{\beta_n} \big\lvert B_{\beta_n} \big\rvert^2 
\delta(\omega-\omega_{\beta_n})
\end{equation}
which we consider within wide band approximation.
In the simulations all inter-molecule hopping parameters are taken $0.01$~eV,
dephasing rates are assumed to be $0.02$~eV.

Figure~\ref{fig4} compares experimental data (panel a) with results of simulation (panel b).
Simulations are performed under bias $V_{sd}=0.02$~eV and external field
illumination at wavelength $\lambda=380$~nm.
In both graphs dark current demonstrates exponential dependence on thickness of molecular layer
characteristic for off-resonant tunneling through wide barrier. Photocurrent shows
insensitivity to barrier width, which is characteristic of hopping transport regime.
We note that the inverse Arrhenius behavior for the photocurrent evident in Figure~\ref{fig3}a indicates that the hopping is  activationless~\cite{RLM7}.

In summary, we presented combined experimental and theoretical study of response
of nitroazobenzene molecular junctions to external illumination and applied bias. 
Experimentally observed characteristic behavior of dark and photo-induced currents was 
modeled within new diagrammatic technique for the Hubbard NEGF.
Being a nonequilibrum atomic limit tool (i.e. formulation employing many-body states 
of isolated molecule as a basis) the Hubbard NEGF readily allows incorporation of the results 
of quantum chemistry simulations into transport behavior. 
We used first principles simulations to model photocurrent behavior and propose mechanisms
behind the observed junction responses. In particular, temperature dependence of the currents
is explained by smearing Fermi-Dirac distribution in the contacts with temperature increase,
which results in increase of dark current due to shifting electron tunneling energies closer
to molecular resonances and decrease of photo-induced current  due to diminished population
in the ground state. 
We note that scattering within molecular layer is an additional factor reducing photocurrent,
which was not included in the theoretical model. 
Also, exponential decrease of dark current with molecular layer thickness
was explained as a manifestation of (off-resonant) tunneling through a barrier,
while insensitivity of photo-induced current on barrier thickness was identified with
hopping regime of transport. 

Further development of the Hubbard NEGF theory, formulation of universal Hubbard NEGF
code for multi-state considerations, and application it to simulation of signals beyond
fluxes are goals of future research.

\begin{acknowledgement}
M.G. and K.M. gratefully acknowledge support by the DOE BES (grant DE-SC0018201).
R.L.M. and A.M.N acknowledge support from the Natural Science and Engineering Research Council of Canada 
and Alberta Innovates. 
\end{acknowledgement}

\begin{suppinfo}
Supporting information contains details on first principle (TD)DFT simulations
and short description of the Hubbard NEGF employed in transport modeling.
\end{suppinfo}


\providecommand{\latin}[1]{#1}
\makeatletter
\providecommand{\doi}
  {\begingroup\let\do\@makeother\dospecials
  \catcode`\{=1 \catcode`\}=2 \doi@aux}
\providecommand{\doi@aux}[1]{\endgroup\texttt{#1}}
\makeatother
\providecommand*\mcitethebibliography{\thebibliography}
\csname @ifundefined\endcsname{endmcitethebibliography}
  {\let\endmcitethebibliography\endthebibliography}{}

\end{document}